# Atomic structure of initial nucleation layer in hexagonal perovskite BaRuO$_3$ thin films

Sangmoon Yoon[1], Jong Mok Ok[1], Sang A Lee[2,3], Jegon Lee[2], Andrew R. Lupini[4], Woo Seok Choi[2]*, and Ho Nyung Lee[1]*

[1]Materials Science and Technology Division, Oak Ridge National Laboratory, Oak Ridge, TN 37831, U.S.A.

[2]Department of Physics, Sungkyunkwan University, Suwon 16419, Korea

[3]Department of Physics, Pukyong National University, Busan 48513, Korea

[4]Center for Nanophase Materials Sciences, Oak Ridge National Laboratory, Oak Ridge, TN 37831, U.S.A.

* e-mail: choiws@skku.edu, hnlee@ornl.gov

# Abstract

Hexagonal perovskites are an attractive group of materials due to their various polymorph phases and rich structure-property relationships. BaRuO$_3$ (BRO) is a prototypical hexagonal perovskite, in which the electromagnetic properties are significantly modified depending on its atomic structure. Whereas thin-film epitaxy would vastly expand the application of hexagonal perovskites by epitaxially stabilizing various metastable polymorphs, the atomic structure of epitaxial hexagonal perovskites, especially at the initial growth stage, has rarely been investigated. In this study, we show that an intriguing nucleation behavior takes place during the initial stabilization of a hexagonal perovskite 9R BaRuO$_3$ (BRO) thin film on a (111) SrTiO$_3$ (STO) substrate. We use high-resolution high-angle annular dark field scanning transmission electron microscopy in combination with geometrical phase analysis to understand the local strain relaxation behavior. We find that nano-scale strained layers, composed of different RuO$_6$ octahedral stacking, are initially formed at the interface, followed by a relaxed single crystal 9R BRO thin film. Within the interface layer, hexagonal BROs are nucleated on the STO (111) substrate by both corner- and face-sharing. More interestingly, we find that the boundaries between the differently-stacked nucleation layers, i.e. heterostructural boundaries facilitates strain relaxation, in addition to the formation of conventional misfit dislocations evolving from homostructural boundaries. Our observations reveal an important underlying mechanism to understand the thin-film epitaxy and strain accommodation in hexagonal perovskites.

Hexagonal perovskites refer to materials with the $ABO_3$ chemical formula, in which the size difference between the A and B cations is large, i.e. the tolerance factor is larger than unity.[1] In these structures, the structural unit, i.e., the $BO_6$ octahedron, share either a face or a corner with a neighboring octahedron to release the internal strain induced by the large ionic size of the A cations.[1] This leads to a variety of polymorph structures, which can be conventionally categorized by the stacking sequence of the $AO_3$ layers (see Fig. 1(a)). For example, the cubic perovskite can be considered as the three-layer cubic (3C) structure within the hexagonal framework, because the $AO_3$ planes are connected by corner-sharing with a repeated stacking pattern after three octahedra. The four-layer hexagonal (4H) structure consists of dimers of face-sharing $BO_6$ octahedra, connected to its neighbors along the *c*-axis by cubic-perovskite-like corner-sharing, with a repeated stacking pattern after four octahedra. The six-layer hexagonal (6H) structure consists of a face-sharing dimer and a single octahedron, connected alternatingly by corner-sharing, with a repeated stacking pattern after six octahedra. The nine-layer rhombohedral (9R) structure consists of trimers of face-sharing $BO_6$ octahedra, connected each other by corner-sharing, with a repeated stacking pattern after nine octahedra. Hexagonal perovskites can possess other varieties of structures depending on their stacking sequence (either by face sharing or corner sharing of the structural units).[2] In this sense, hexagonal perovskites serve as a platform for the investigation of a myriads of structure-property relationships.[3-6]

In particular, Ba-based $ABO_3$ oxides, such as $BaRuO_3$ (BRO) and $BaIrO_3$ (BIO), have attracted considerable attention as prototypical hexagonal perovskites, as each polymorph exhibits unique electronic and magnetic properties.[7,8] In the case of the BRO system, the cubic phase (3C) ($a = b = c = 4.005$ Å, space group 221 ($Pm\bar{3}m$))[9] was found to be a ferromagnetic metal with a transition temperature $T_c \approx 60$ K, whereas the 4H phase ($a = b = 5.729$ Å, $c = 9.500$ Å, space group 194 (P63/mmc))[10] and 9R phase ($a = b = 5.755$ Å, $c = 21.621$ Å, space group 166 ($R\bar{3}m$))[11,12] are paramagnetic.[8] Electronically, the 4H phase was metallic down to the base temperature while the 9R phase showed a metal to semiconductor transition at 150 K with a pseudogap opening.[13-15] On the other hand, the 6H BIO phase ($a$ = 5.748 Å, $b$ = 9.939 Å, $c$ = 14.358 Å, space group 15 (C2/c))[16] was shown to be a paramagnetic metal, but the 9R BIO phase ($a$ = 10.005 Å, $b$ = 5.754 Å, $c$ = 15.184 Å, space group 12 (C2/m))[16] undergoes a ferromagnetic transition at $T_c \approx$ 180 K accompanied by a semiconductor to insulator transition.[7] The physical properties of the hexagonal perovskite BaBO$_3$ (B = Ru, Ir) have been extensively investigated over the last decades, but the exact electromagnetic ground states are still elusive, possibly owing to the strongly correlated nature within the hexagonal structural framework. For example, the electronic transitions in both 9R BRO and BIO phases are thought to be related to the formation of a charge density wave (CDW)[13-15, 17], and the interaction with molecular orbital states have been suggested as the possible underlying mechanism.[8, 18] Nevertheless, it is evident that the dissimilar portion of the orbital overlap between the B-site ions within the face-sharing dimers or trimers will greatly affect the physical properties of Ba-based hexagonal perovskites.

As the hexagonal polymorphs are formed by alternate stacking of the hexagonal AO$_3$ layer along the *c*-axis, the hexagonal surface symmetry of the (111) cubic-perovskite substrates provides good compatibility for the epitaxial thin film growth. It has been reported that polycrystalline hexagonal perovskite thin films containing various polymorph phases are grown with smoother surface structure on the (111) SrTiO$_3$ (STO) substrate compared to those on the (001) or (011) substrates.[19, 20] While some of the polycrystalline hexagonal perovskite thin films have been studied using X-ray diffraction and transmission electron microscopy (TEM)[21, 22], the microstructure of the single-phase thin film has not yet been investigated at the atomic scale, largely owing to the lack of high-quality epitaxial thin films. The formation mechanism of the high-quality hexagonal perovskites at the initial stage of growth on the (111) cubic-perovskite substrates would provide crucial information for stabilizing and further strain-engineering the metastable hexagonal perovskite phases through thin-film epitaxy.

In this work, we investigated the interfacial atomic structures of a 9R BRO thin film grown on a (111) STO substrate using high-resolution high-angle annular dark field scanning transmission electron microscopy (HAADF STEM) with geometrical phase analysis (GPA). 9R BRO was chosen as it has the lowest formation energy among the BRO polymorphs,[9] and therefore, grows in single phase. We observed that a thin (4 – 6 nm) intermediate layer was formed between the bulk-like 9R BRO thin film and the (111) STO substrate. Interestingly, both the face- as well as the corner-shared layers were observed to coexist in the nucleation layer. Furthermore, we find that the boundaries between the distinctively-stacked nucleation layers act as strain relaxation sites in addition to the conventional misfit dislocations.

Single-phase 9R BRO thin films (~30 nm) were epitaxially grown on the STO (111) substrate using pulsed laser deposition at the growth temperature of 750 °C and oxygen partial pressure of 300 mTorr. Cross-sectional STEM specimens were prepared via ion milling at $LN_2$ temperature after conventional mechanical polishing. High-resolution HAADF STEM experiments were carried out using a Nion UltraSTEM100 operated at 100 kV. The microscope was equipped with a cold field emission gun and a corrector of third- and fifth-order aberration for sub-angstrom resolution. The collection inner and outer half-angles for HAADF STEM were 65 and 240 mrad, respectively. Noise and amorphized surface background arising in the high-resolution HAADF STEM images were reduced with a band-pass filter. GPA and multislice electron-scattering simulation were performed using ER-C GPA[23] and QSTEM[24] code, respectively.

The global lattice structure of an epitaxial 9R BRO thin film on a (111) STO substrate is evidenced via X-ray diffraction $\theta$-$2\theta$ scan (Fig. S1). The thin film is indeed composed of pure 9R phase without any impurity or other polymorph phase, which has not been reported so far. Fig. S2 presents a low-magnification HAADF STEM image, further demonstrating that the 9R BRO thin film is uniformly grown on (111) STO without any impurities. Fig. 1(b) shows a high-magnification HAADF STEM image of 9R BRO viewed along the [11$\bar{2}$0] zone axis. The lattice parameters measured in HAADF STEM are $a$ = 5.75 Å and $c$ = 21.59 Å, consistent with the XRD result ($c$ = 21.60 Å) and the bulk values.

Owing to the large lattice mismatch nominally imposing ~4% tensile strain, the film is fully relaxed except for the interface region. Large and small bright dots in the HAADF STEM indicate Ba and Ru atomic columns, respectively, indicating that the STEM image viewed along the [11$\bar{2}$0] zone axis can be used to directly distinguish the atomic structures of the hexagonal polymorph. The discernability of hexagonal BRO polymorphs in the HAADF STEM image viewed along the [11$\bar{2}$0] zone axis is further demonstrated in Fig. 1(c) using simulated HAADF STEM images. Note that O atomic columns are not detected in the HAADF STEM images, since the scattering intensity in the HAADF STEM image depends strongly upon atomic number. The local connection between the $RuO_6$ octahedra in hexagonal perovskites, i.e. face- and corner-sharing, can be analyzed by HAADF STEM. The face-sharing of $RuO_6$ octahedra corresponds to Ru atomic columns aligned along the *c*-axis, while the corner-sharing corresponds to Ru atomic columns lying aslant from that axis.

Because the growth of epitaxial thin film is predominantly determined at the nucleation stage, it is critical to characterize the atomic structures of the interface between the thin film and the substrate.[25,26] Fig. 2(a) shows a typical high-magnification HAADF STEM image for the interfacial region between a 9R BRO thin film and a (111) STO substrate. The image shows that an intermediate layer (denoted by white dashed lines in Fig. 2(a)), where the atomic structure is different from that of nominal 9R BRO, is formed between the bulk film and the substrate. To systematically characterize the atomic structures of the intermediate layer, the interfacial region denoted by the parenthesis in Fig. 2(a) is further magnified in Fig. 2(b). The high-magnification image reveals two distinct features of the intermediate layer. First, the intermediate layer is composed of dissimilar octahedral face-sharing chain structures other than the nominal trimer, including single octahedron, dimer, and tetramer. The octahedral chains even longer than the tetramer were sometimes observed within the intermediate layer (see Fig. S3 in the Supplementary Information). Second, the irregular atomic configurations are frequently extended from the film/surface interface to about 4 – 6 nm above into the intermediate layers (vertical red arrows). Those irregular atomic structures form the type-I relaxation sites, which will be discussed below.

Additionally, the image directly demonstrates that the initial $RuO_6$ octahedra in contact with $TiO_6$ octahedra of the STO substrate are both face-sharing as well as corner-sharing. The nucleation layers face- and corner-sharing with the substrate are denoted by the blue and red solid lines in Fig. 2(b), respectively. Note that the corner-shared nucleation layer occupies the majority of the interface connection, while the face-shared nucleation layers are formed as islands with ~10 nm size.

To further examine the strain state of the intermediate layer and 9R BRO bulk film, GPA was applied to the HAADF STEM images. Fig. 2(c) shows an in-plane strain ($\varepsilon_{xx}$) map of Fig. 2(a) This strain map demonstrates that 9R BRO bulk film is fully relaxed while the intermediate layer is in the partially-strained state though the strain relaxes at several specific points. This further suggests that the formation of non-bulk octahedral chains such as dimers or tetramers in the intermediate layer is related to the metastability of BRO polymorphs in the partially strained state. In addition, the direct comparison of the strain map and the HAADF STEM image reveals that the strain relaxation sites can be classified into two distinctive types depending on the characteristics of the BRO nucleation layer encountered at the relaxation core; type-I and II relaxation sites are denoted in Fig. 2 by the red and blue arrows, respectively. The characteristics of each relaxation sites will be discussed in the following paragraph. The atomic structures denoted by the red arrows in Fig. 2(b) show that the type-I relaxation sites with homostructural boundaries correspond to conventional misfit dislocations formed by a large lattice mismatch. It seems that an atomic-scale core is formed between the face-shared nucleation layers, and the dislocation structures extend from the interfacial core to about 4 ~ 6 nm above in distance. The strain map in Fig. 2(b) additionally shows that the epitaxial strain is relaxed in the type-I relaxation sites via a two-step process. The strain is partly relaxed at the atomic-scale core within the nucleation layer and finally fully relaxed at the other end-point of the dislocation structure. These type-I relaxation sites are important because they define the thickness of the intermediate (strain relaxation) layer. Meanwhile, the atomic structures denoted by the blue arrows in Fig. 2(b) show that the type-II relaxation sites are the line segments where the face-shared and corner-shared nucleation layers meet. In other words, the type-II relaxation sites are the stacking disorder boundary (heterostructural boundary) occurring at the nucleation layer of the hexagonal 9R BRO thin film. The core structure was blurred and not clearly resolved in the HAADF STEM images, suggesting that the boundaries are not perfectly aligned along the [1$\bar{1}$0] STO ∥ [11$\bar{2}$0] BRO direction. The sites where the epitaxial strain is relaxed are presented in the schematic of the nucleation layer in Fig. 3.

The atomic-scale characterization of the 9R BRO thin film grown on the (111) STO substrate provides important insights in the thin film epitaxy of hexagonal perovskites. First, the (111) cubic-perovskite substrate is verified as an ideal choice for hexagonal perovskite growth in terms of chemical compatibility and lattice symmetry; a pure 9R BRO epitaxial film was formed without any impurities or polymorphs. Second, nano-scale intermediate layers composed of non-bulk octahedral chains are observed at the interface, which is attributed to the partially-strained state imposed on the hexagonal BRO. This observation suggests that various metastable BRO polymorphs with similar hexagonal structural framework might be stabilized in the thin-film form owing to their small formation energy difference, of which the growth could be further controlled by the epitaxial strain. Moreover, our results reveal that the epitaxial strain is relaxed at the boundaries where the face-shared and corner-shared nucleation layers meet, as well as the conventional misfit dislocations. This is a clear distinction between strain relaxations in cubic and hexagonal perovskites. While the lattice mismatch and strain susceptibility are the dominant factors to be considered for strain-engineering of cubic perovskite, the stacking of nucleation layers also needs to be considered and controlled for hexagonal perovskite to be strained.

In summary, we investigated the atomic structure of a hexagonal 9R BRO thin film grown on a (111) STO substrate using high-resolution HAADF STEM imaging with GPA strain mapping. Nanoscale intermediate layers composed of different $RuO_6$ octahedral stacking were formed between the relaxed 9R BRO bulk film and the STO substrate. The formation of the intermediate layers can be attributed to the partial strain imposed on the interfacial region. The BRO thin film was observed to be nucleated on the (111) STO substrate by face-sharing as well as corner-sharing. The stacking disorder boundaries were found to act as the relaxation sites in addition to the conventional misfit dislocations. This study demonstrates that thin-film epitaxy can be used as a promising way to stabilize metastable or artificial BRO polymorphs. Furthermore, our study suggests that the control of nucleation stacking is of critical importance for growing epitaxial thin films of hexagonal perovskites.

## Acknowledgements

This work was supported by the U.S. Department of Energy, Office of Science, Basic Energy Sciences, Materials Sciences and Engineering Division. The electron microscopy work was conducted at the Center for Nanophase Materials Sciences, which is a DOE Office of Science User Facility. W.S.C., J.L., and S.A.L were supported by the Basic Science Research Programs through the National Research Foundation of Korea (NRF-2019R1A2B5B02004546 (W.S.C. and J.L.) and NRF-2019R1A2C1005267 (S.A.L)).

# Figures

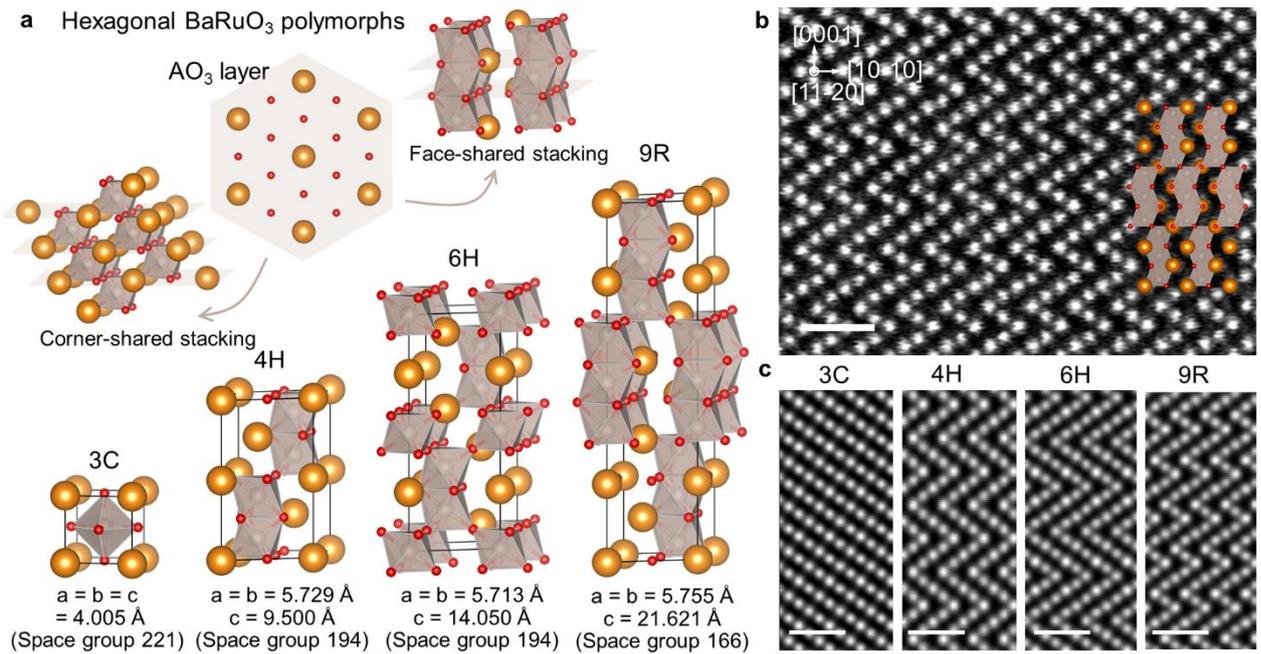

**Figure 1.** Hexagonal perovskite polymorphs of BaRuO$_3$ (BRO) and HAADF STEM images. (a) Hexagonal BRO polymorphs; three-layer cubic (3C), four-layer hexagonal (4H), six-layer hexagonal (6H), and nine-layer rhombohedral (9R). (b) High-magnification HAADF STEM image of a 9R BRO thin film viewed along the BRO [11$\bar{2}$0] zone axis. (c) Simulated HAADF STEM images of 3C, 4H, 6H, and 9R BRO viewed along the BRO [11$\bar{2}$0] zone axis. The scale bar indicates 1 nm.

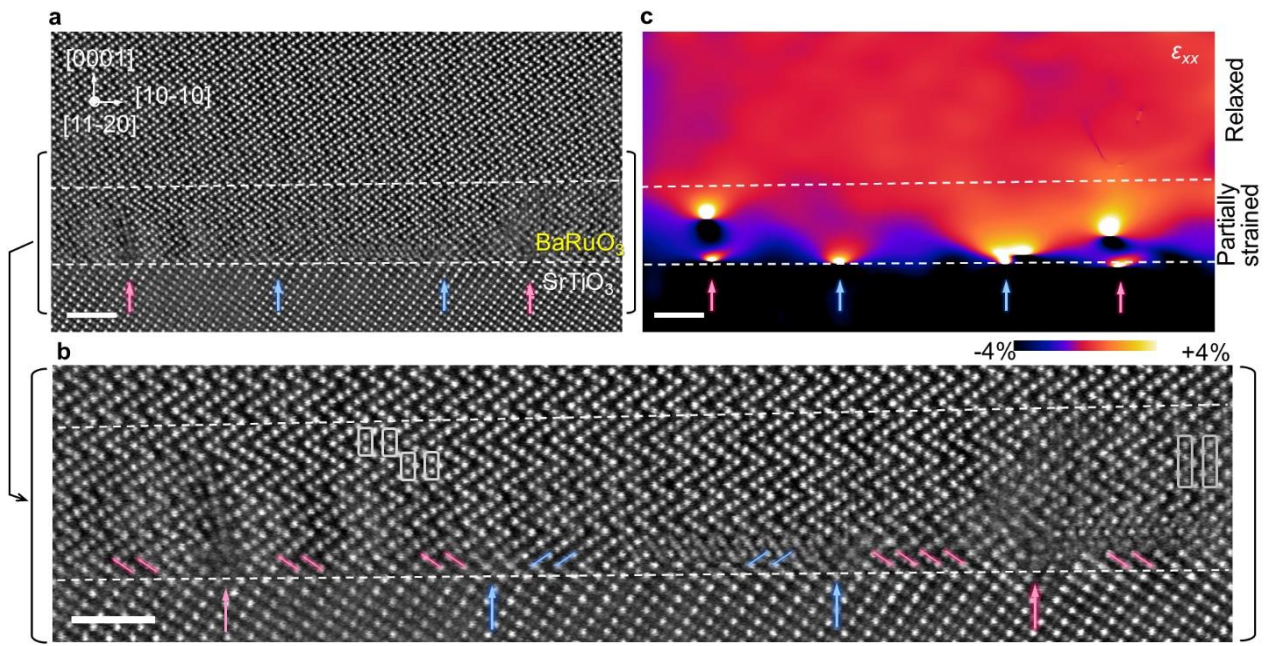

**Figure 2.** Microstructure and strain map of the 9R BRO/(111) STO interface. (a) High-resolution HAADF STEM image of a 9R BRO/(111) STO interface viewed along the BRO [11$\bar{2}$0] zone axis. The intermediate layer on (111) STO is marked by the white dotted lines. (b) Magnified HAADF STEM image of the interfacial region marked by the black parenthesis in Fig. 2(a). The corner- and face-shared stackings at the nucleation layer are highlighted by the red and blue solid lines, respectively. RuO$_6$ dimer and tetramer formed in the intermediate layer are denoted by the bright and dark grey rectangles, respectively. (c) In-plane strain map ($\varepsilon_{xx}$) of the 9R BRO/(111) STO interface obtained via GPA strain analysis. The type-I and II relaxation sites are denoted by the blue and red arrows in the images, respectively. The scale bar indicates 2 nm.

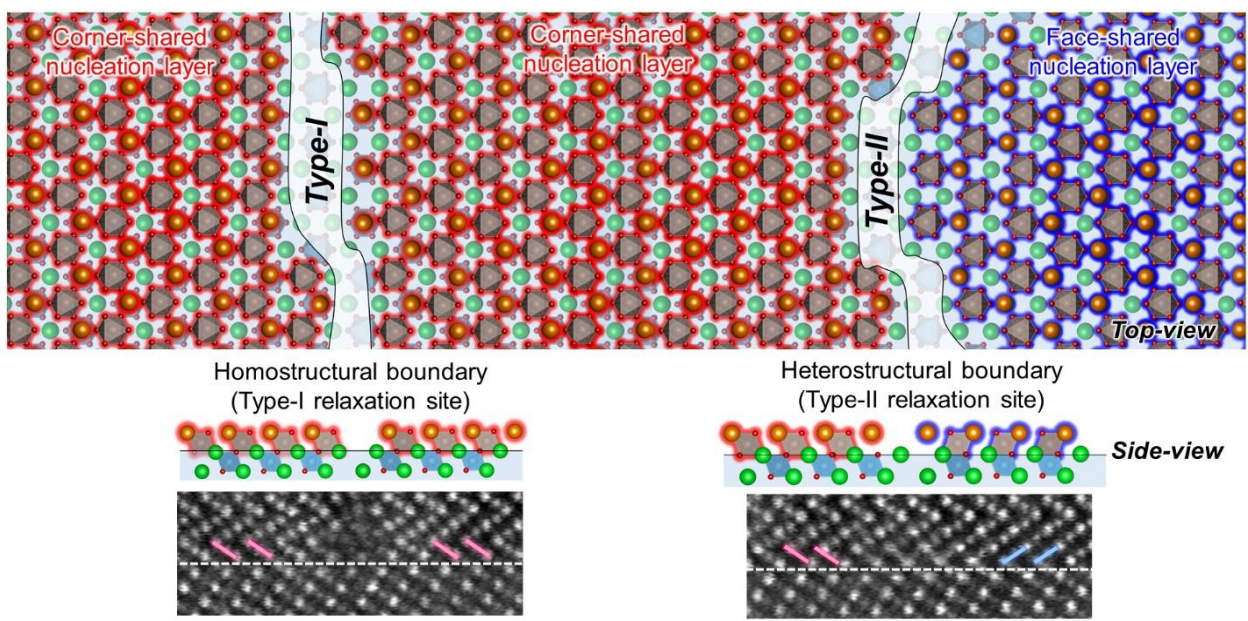

**Figure 3.** Top- and side-view schematics for the BRO nucleation layer. The bright blue and grey octahedra indicate $TiO_6$ and $RuO_6$ octahedra, while green and orange spheres indicate Sr and Ba atoms, respectively. The red- and blue-glowing layers indicate the corner- and face-shared $RuO_6$ nucleation layers, respectively. Type-I and II relaxation sites are formed between the face-shared nucleation layer (homostructural boundary) and at the stacking disorder boundary (heterostructural boundary), respectively. Examples of HAADF STEM images for the type-I and II relaxation sites are represented below the schematics, which were extracted from Fig. 2(b).

# Atomic structure of initial nucleation layer in hexagonal perovskite BaRuO$_3$ thin films

Sangmoon Yoon[1], Jong Mok Ok[1], Sang A Lee[2,3], Jegon Lee[2], Andrew R. Lupini[4], Woo Seok Choi[2]*, and Ho Nyung Lee[1]*

[1]Materials Science and Technology Division, Oak Ridge National Laboratory, Oak Ridge, TN 37831, U.S.A.

[2]Department of Physics, Sungkyunkwan University, Suwon 16419, Korea

[3]Department of Physics, Pukyong National University, Busan 48513, Korea

[4]Center for Nanophase Materials Sciences, Oak Ridge National Laboratory, Oak Ridge, TN 37831, U.S.A.

* e-mail: choiws@skku.edu, hnlee@ornl.gov

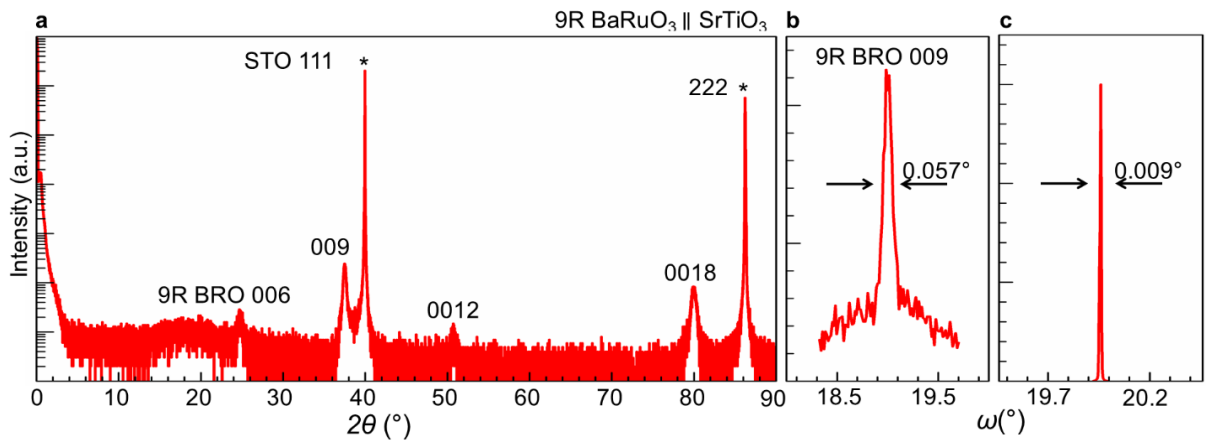

**Figure S1**. (a) X-ray diffraction (XRD) θ-2θ pattern of a 9R-BRO thin film on a (111) STO substrate. The rocking curve scans for the (b) thin film and (c) substrate, indicating the high quality of the thin film.

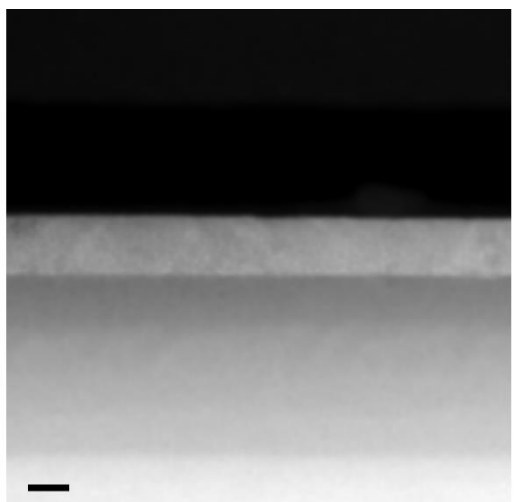

**Figure S2.** A low-magnification HAADF STEM image of a 9R-BRO thin film. The low-magnification HAADF STEM image verifies that the overall 9R BRO thin film is uniformly grown on a (111) STO substrate without any impurities. The scale bar indicates 20 nm.

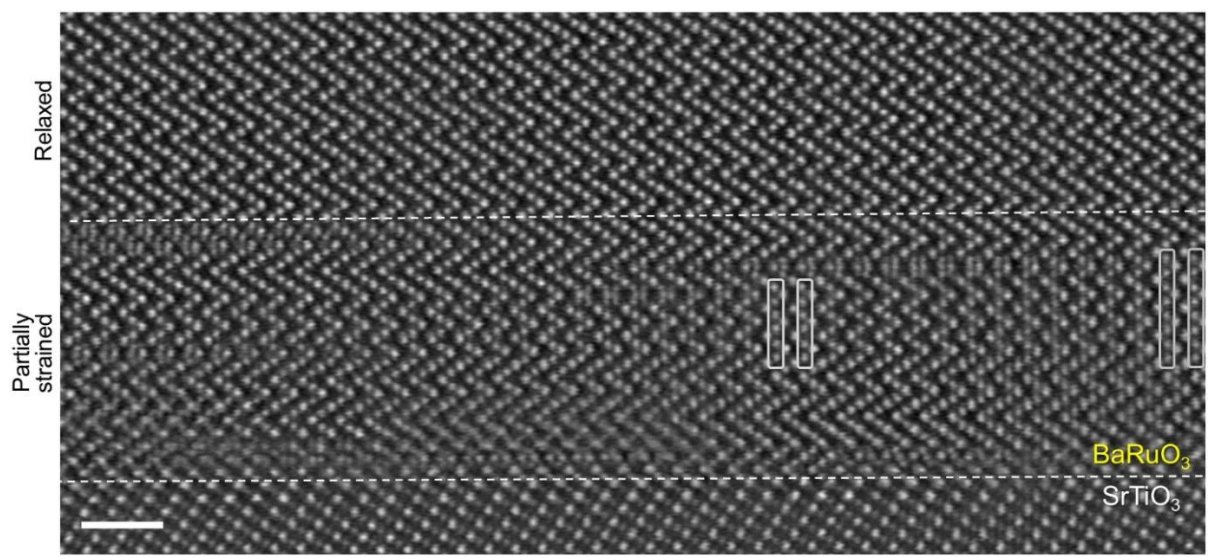

**Figure S3.** High-resolution HAADF STEM image of a 9R-BRO thin film, where long RuO octahedral chains are formed in the interfacial region. The HADDF STEM image is viewed along the [1-10]STO∥[11$\bar{2}$0]BRO zone axis. The scale bar indicates 2 nm. The 9R-BRO/STO interface and the border between partially strained and fully relaxed regions are marked by the white dotted lines. The atomic chains consisting of six and eight $RuO_6$ octahedra formed near the interface are highlighted by the bright and dark grey rectangles, respectively. This image shows that unusually long $RuO_6$ octahedral chains are sometimes formed at the partially strained (interfacial) region.

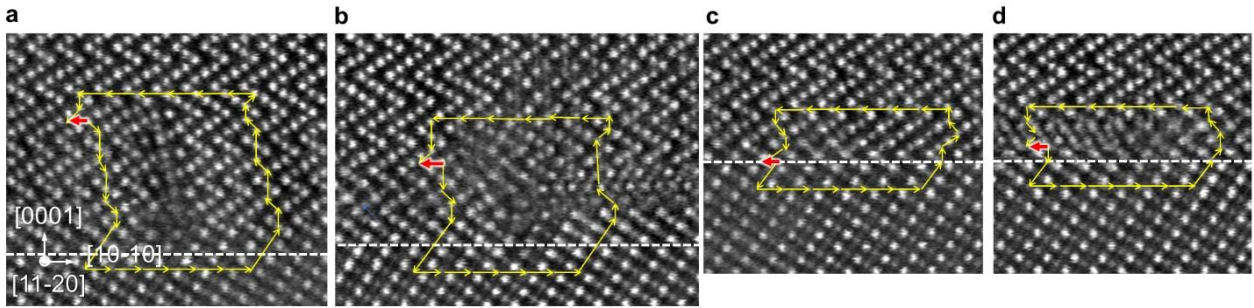

**Figure S4.** Analysis of Burgers circuit for type-I and II relaxation sites; (a)-(b) type-I and (c)-(d) type-II relaxation sites. The Burgers circuits are denoted by yellow arrows and the resultant Burgers vectors are denoted by red arrows. The 9R-BRO/STO interface are marked by the white dotted lines in the images. Type-II as well as type-I relaxation sties commonly have the Burger vectors along the BRO [10$\bar{1}$0] direction. In other words, the type-II relaxation sites are also misfit dislocations, but they are the unconventional ones formed at staking disorder boundaries.